\documentclass[
reprint,
floatfix,
aps,
prl,
amsmath,
twocolumn,
superscriptaddress,
amssymb,
tightenlines,
groupaddress,
footinbib
]{revtex4-1}

\usepackage{amsfonts,amssymb}
\usepackage[subnum]{cases}
\usepackage{mathrsfs}
\usepackage{amsmath}
\usepackage[none]{hyphenat}
\usepackage{hyperref} 
\usepackage{bm} 
\usepackage{natbib}
\usepackage{soul} 
\usepackage{color}
\usepackage{longtable}
\usepackage{ textcomp }
\usepackage{verbatim}
\usepackage{fancyhdr} 
\usepackage{lastpage} 
\usepackage{extramarks} 
\usepackage{graphicx} 
\usepackage{lipsum} 
\usepackage{amssymb}
\usepackage{slashed}
\usepackage{tikz}
\usepackage{comment}
\usepackage{natbib}
\usepackage[caption=false,listofformat=parens, subrefformat=parens]{subfig}
\usepackage{marvosym}
\usepackage{array}

\usepackage{amsthm} 

\newcommand{\abs}[1]{\left| #1 \right|} 
\newcommand{\dd}[2]{\frac{d^2 #1}{d #2^2}} 
 
\newcommand{\ket}[1]{\left| #1 \right>} 
\newcommand{\bra}[1]{\left< #1 \right|} 
\let\baraccent=\= 
\renewcommand{\=}[1]{\stackrel{#1}{=}} 

\theoremstyle{definition}

\theoremstyle{remark}

\newcolumntype{C}[1]{>{\centering\let\newline\\\arraybackslash\hspace{0pt}}m{#1}}

\newcommand{\la}{\left <}
\newcommand{\ra}{\right >}

\def\ar{\rho_{xx}/\rho_{yy}}

\def \Rxx{R_{xx}}
\def \Ryy{R_{yy}}

\newcommand{\req}[1]{Eq.\,(\ref{#1})}

\newcommand{\rfig}[1]{Fig.\,\ref{#1}}

\newcommand{\rref}[1]{Ref.\,\onlinecite{#1}}

\usepackage{physics}
\usepackage{CJK}

\begin{document}
\begin{CJK*}{UTF8}{gbsn}
\title{Resistivity anisotropy of quantum Hall stripe phases}
\author{M.~Sammon}
\email[Corresponding author: ]{sammo017@umn.edu}
\author{X.~Fu}
\author{Yi~Huang~(黄奕)}
\author{M.~A.~Zudov}
\author{B.~I.~Shklovskii} 
\affiliation{School of Physics and Astronomy, University of Minnesota, Minneapolis, MN 55455, USA}
\author{G.\,C. Gardner}
\affiliation{Microsoft Quantum Lab Purdue, Purdue University, West Lafayette, Indiana 47907, USA}
\affiliation{Birck Nanotechnology Center, Purdue University, West Lafayette, Indiana 47907, USA}
\author{J.\,D. Watson}
\altaffiliation[Present address: ]{Microsoft Station-Q at Delft University of Technology, 2600 GA Delft, The Netherlands}
\affiliation{Birck Nanotechnology Center, Purdue University, West Lafayette, Indiana 47907, USA}
\affiliation{Department of Physics and Astronomy, Purdue University, West Lafayette, Indiana 47907, USA}
\author{M.\,J. Manfra}
\affiliation{Microsoft Quantum Lab Purdue, Purdue University, West Lafayette, Indiana 47907, USA}
\affiliation{Birck Nanotechnology Center, Purdue University, West Lafayette, Indiana 47907, USA}
\affiliation{Department of Physics and Astronomy, Purdue University, West Lafayette, Indiana 47907, USA}
\affiliation{School of Electrical and Computer Engineering and School of Materials Engineering, Purdue University, West Lafayette, Indiana 47907, USA}
\author{K. W. Baldwin}
\affiliation{Department of Electrical Engineering, Princeton University, Princeton, New Jersey 08544, USA}
\author{L. N. Pfeiffer}
\affiliation{Department of Electrical Engineering, Princeton University, Princeton, New Jersey 08544, USA}
\author{K. W. West}
\affiliation{Department of Electrical Engineering, Princeton University, Princeton, New Jersey 08544, USA}
\received{\today}

\begin{abstract}
Quantum Hall stripe phases near half-integer filling factors $\nu \ge 9/2$ were predicted by Hartree-Fock (HF) theory and confirmed by discoveries of giant resistance anisotropies in high-mobility two-dimensional electron gases. 
A theory of such anisotropy was proposed by MacDonald and Fisher, although they used parameters whose dependencies on the filling factor, electron density, and mobility remained unspecified. 
Here, we fill this void by calculating the hard-to-easy resistivity ratio as a function of these three variables. 
Quantitative comparison with experiment yields very good agreement which we view as evidence for the ``plain vanilla'' smectic stripe HF phases.
\end{abstract}
\maketitle
\end{CJK*}

\begin{figure}[t]
\includegraphics[width=\linewidth]{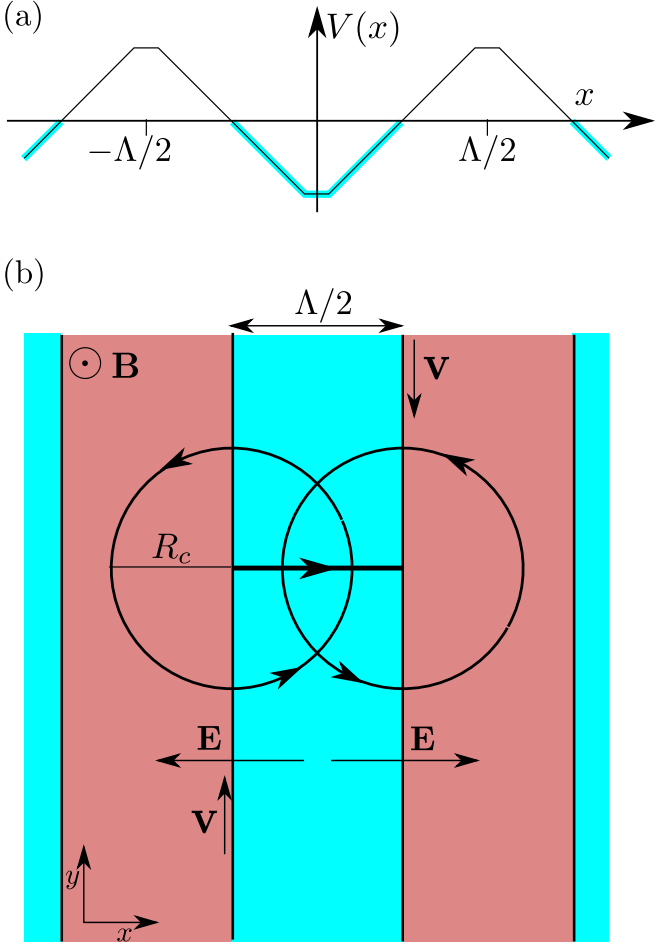}
\vspace{-0.1 in}
\caption{(a) Schematic drawing of the single particle self-consistent potential energy $V(x)$ which is responsible for stripe formation due to approximate boxlike electron-electron interaction \citep{fogler:1996}.
The sloped regions of $V(x)$ determine the internal electric field $E$. 
(b) Schematic of transport in the stripe phase. Electrons on the stripe edges (shown in blue) drift in electric fields $E$ with velocity $v$ in the $\pm y$-direction. 
They are scattered to an adjacent stripe edge by background charged impurities at a rate $2/\tau_B$, as illustrated by thick horizontal arrow.}
\label{fig:stripes}
\vspace{-0.2 in}
\end{figure}

Quantum Hall stripe phases near half-integer filling factors $\nu \ge 9/2$ were predicted for spin-split Landau levels (LLs) by the Hartree-Fock (HF) theory~\citep{koulakov:1996,fogler:1996,moessner:1996}.
At exactly half-integer filling factor $\nu$, these phases consist of alternating stripes with filling factors $\nu - 1/2$ and $\nu + 1/2$, both with the width $\Lambda/2 \simeq 1.4 R_c$~\citep{koulakov:1996,fogler:1996,wexler:2001,rezayi:1999}, where $R_c$ is the cyclotron radius (see Fig.~\ref{fig:stripes}). 
These stripes are formed due to the repulsive box-like screened interaction potential of electrons with ring-like wave functions in high LLs. 
Such a potential leads to an energy gain when electrons occupy the nearest states of the same stripe and avoid interacting with electrons in neighboring stripes. 
The self-consistent HF theory is valid at LL indices $N\gg 1$, when $R_c =l_B(2N+1)^{1/2} \gg l_B$. 
Here, $l_B = (c\hbar/eB)^{1/2}$ is the magnetic length, which is a measure of quantum fluctuations of an electron's cyclotron orbit center. 
It was shown \citep{fogler:1996,wexler:2001,rezayi:1999} that quantum fluctuations play little role even when $N=2$, so that stripes should determine ground states for all half-integer $\nu \ge 9/2$.  

Quantum Hall stripes were confirmed by discoveries of dramatic resistance anisotropies in high-mobility two-dimensional electron gases (2DEGs) hosted in GaAs/AlGaAs heterostructures at $\nu = 9/2, 11/2, 13/2, ...$~\citep{lilly:1999a,du:1999}.
The preferred direction of the stripes (symmetry breaking) was found to be imposed by a potential related to GaAs crystal orientation, whose origin is not understood even now. 

MacDonald and Fisher (MF) proposed a theory of the stripe phase conductivity \citep{macdonald:2000}. 
They assumed that stripes form a smectic state, pinned by disorder, and used an analogy between stripe edges and edge states in a confined 2DEG (see Fig.~\ref{fig:stripes}). 
At half-integer filling factors $\nu \geq 9/2$ this theory leads to the resistivity ratio
\begin{equation}\label{eq:alphaMF}
\frac{\rho_{xx}}{\rho_{yy}} = \left(\frac{v\tau_B}{\Lambda}\right)^2 \gg 1, 
\end{equation}
where $\Lambda \simeq 2.8 R_c$ is the stripe period, $v$ is the drift velocity of electrons on stripes edges (see Fig.~\ref{fig:stripes}), and $\tau_B$ is the time of an electron scattering to a neighboring stripe edge. Let us interpret Eq.~\eqref{eq:alphaMF}. 
An electron drifts for a time $\tau_B/2$ until it is scattered to one of the adjacent edges. 
Thus, we can define two electron diffusion constants
\begin{align}\label{eq:D_x}
	&D_{xx}=\frac{1}{2}\frac{(\Lambda/2)^2}{\tau_B/2}=\frac{\Lambda^2}{4\tau_B},\\
	&D_{yy}=\frac{1}{2}\frac{(v \tau_B/2)^2}{\tau_B/2}=\frac{v^2\tau_B}{4}.  \label{eq:D_y}
\end{align}
Here, we have used the fact that at each time step $\tau_B/2$ an electron on the edge of a stripe randomly moves a distance $v\tau_B/2$ in the $y$-direction, while it hops a distance $\Lambda/2$ in the $x$-direction. Taking the ratio of $D_{yy}$ and $D_{xx}$ and using the Einstein relationship we arrive at Eq.~\eqref{eq:alphaMF}.

In its present form, Eq.~\eqref{eq:alphaMF} does not allow comparison with the experimental data, which we talk about below, as \rref{macdonald:2000} did not specify how $\tau_B$ or $v$ depend on the electron density $n_e$, the mobility $\mu$, and the filling factor $\nu$. 
In this Rapid Communication we calculate $\tau_B$ and $v$, and arrive at the ratio of resistivities which can be directly compared with experimental data,
\begin{equation}\label{eq:alpha}
\frac{\rho_{xx}}{\rho_{yy}}= 
\frac{0.088}{\gamma^2}\left(\frac{\hbar n_e \mu}{\pi^3 e \nu(2N+1)}\right)^2.
\end{equation}
Here, $\nu \geq 9/2$ is either $2N+1/2$ or $2N+3/2$ and $\gamma\equiv\gamma(N_1/N_2)$ is a dimensionless function of concentrations $N_1$ and $N_2$ of unintentional background impurities in the Al$_x$Ga$_{1-x}$As spacer and GaAs quantum well, correspondingly. 
We focus on ultrahigh mobility 2DEG in which the long range potential of remote donors plays a minor role for the momentum relaxation time $\tau$ and $\tau_B$. 
It is easy to see from Eq.~\eqref{eq:alpha} that $\ar \propto  \mu^2 B^4/n_e^2$ at $N \gg 1$. 
We show below that this prediction for high LL agrees with experiment and arrive at $N_1/N_2 \simeq 60$ using $N_1/N_2$ as a single fitting parameter.
\begin{figure}[t]
	\centering
	\includegraphics[width=\linewidth]{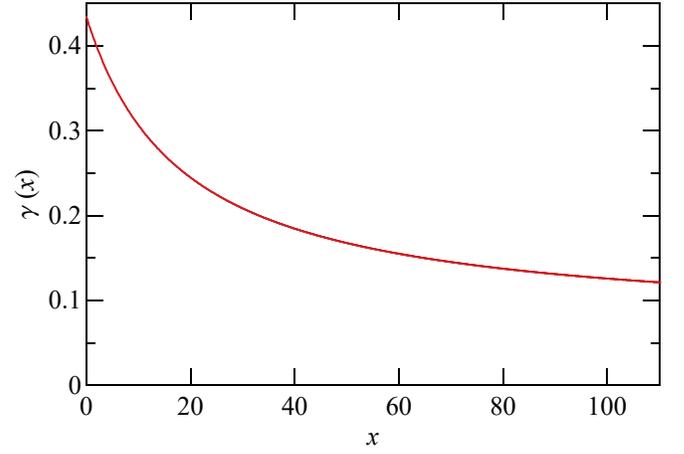}
\vspace{-0.1 in}
\caption{$\gamma(x)$ for electron density $n_e=3\times10^{11}$ cm$^{-2}$ and quantum well width $w=30$ nm.}
\label{fig:gamma}
\vspace{-0.2 in}
\end{figure}
	
Let us now derive our expressions for $v$ and $\tau_B$ which allow the conversion of Eq.~\eqref{eq:alphaMF} into Eq.~\eqref{eq:alpha}. 
The drift velocity is $v = cE/B$, where $B$ is the magnetic field, $eE=|dV/dx|$ at $x=\pm \Lambda/4$ is the internal electric field at the stripe edges, and $V(x)$ is the self-consistent HF potential.
Following MF, we assume that at low temperatures electrons form a smectic pinned by impurities. 
If we use the model of a saw-tooth stripe potential $V(x)$ (Fig.~\ref{fig:stripes}a), based on the simplified box model of electron repulsive potential given by Eq.~(15) of Ref.~\onlinecite{fogler:1996}, we find $eE=\hbar\omega_c/2\pi^2R_c$, where $\omega_c$ is cyclotron frequency. 
Below we use more accurate expression 
\begin{equation}\label{eq:E}
	eE=\beta(r_s) \frac{\hbar\omega_c}{2\pi^2R_c},
\end{equation} 
where $\beta(r_s) $ can be obtained from Eqs.~(48) and (43) of Ref.~\onlinecite{fogler:1996}, $r_s\equiv (\pi n_e a_B^2)^{-1/2}$, $a_B = \hbar^2 \kappa / m^\star e^2$, $m^\star$ is the effective mass, and $\kappa$ is the dielectric constant of GaAs. 
The 2DEGs we consider have $r_s\simeq 1$ and $\beta(1)=0.77$. 
Using \req{eq:E} we find the drift velocity in Eq.~\eqref{eq:alphaMF}
\begin{equation}\label{eq:v_D}
	v= \frac {cE} B = \frac{\beta}{2\pi^2}\frac{v_F}{\sqrt{\nu(2N+1)}}\,,
\end{equation}
where $v_F$ is the Fermi velocity.
Next, we show that 
\begin{equation}\label{eq:tau_star}
	\frac{2}{\tau_B}=\frac{\gamma}{\tau}\frac{g_B}{g_0},
\end{equation} 
where $g_0=m^\star/2\pi\hbar^2$ is the density of states per spin at $B = 0$, and
\begin{equation}\label{eq:DOS}
	g_B=\frac{2}{2\pi l_B^2eE\Lambda}=\frac{2}{hv\Lambda}
\end{equation}
is the modified density of states at the Fermi level of the spin-polarized half-filled LL [see Fig.~\ref{fig:stripes}(a)], defined as the ratio of $(2\pi l_B^2)^{-1}$ and the energy width of a LL $eE\Lambda/2$ \citep{fogler:1996}. 
Apparently, in strong magnetic fields relatively narrow LLs, with $g_B/g_0 \gg 1$, are formed. 
This increases the scattering rate $2/\tau_B$ in comparison with $1/\tau$~\citep{dmitriev:2003} in Eq.~\eqref{eq:tau_star}. 
The dimensionless function $\gamma(N_1/N_2)$ in Eq.~\eqref{eq:tau_star}, shown in Fig.~\ref{fig:gamma}, takes care of relative contributions of the background charged impurities in the spacer and in the quantum well to $2/\tau_B$ and $1/\tau$ (See Supplementary Material). 
Combining Eqs.~\eqref{eq:alphaMF},~\eqref{eq:v_D},~\eqref{eq:tau_star}, and~\eqref{eq:DOS} we arrive at Eq.~\eqref{eq:alpha}.

\begin{figure}[b]
\vspace{-0.1 in}
\includegraphics{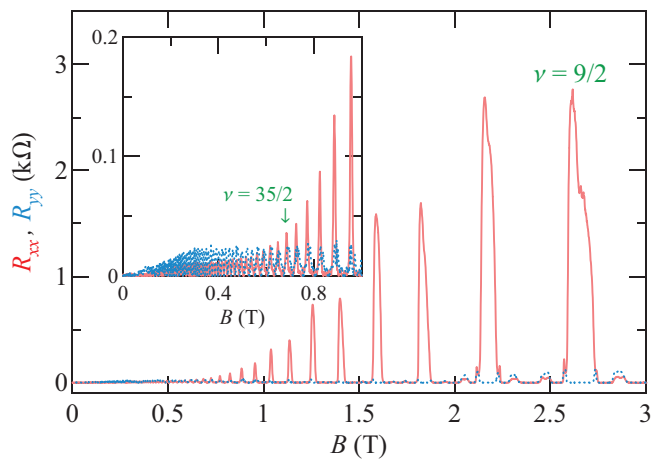}
\vspace{-0.1 in}
\caption{
(color online) 
$R_{xx}$ (solid line) $R_{yy}$ (dotted line) as a function of $B$ measured in sample A at $T \approx 50$ mK.}
\vspace{-0.2 in}
\label{fig:data}
\end{figure}

\begin{table*}
\caption{Sample ID, electron density $n_e$, mobility $\mu$, quantum well width $w$, setback distance $d$.}
\begin{ruledtabular}
\begin{tabular}{c c c c c} 
Sample ID & $n_e$ ($10^{11}$ cm$^{-2}$) & $\mu$ ($10^6$ cm$^2$/Vs) & $w$ (nm) & $d$ (nm)\\
A & 2.9 &  28 & 30   & 75  \\ 
B & 2.8 &  16 & 30   & 75  \\ 
C & 3.0 &  16 & 30   & 80  \\ 
\end{tabular}
\end{ruledtabular}
\vspace{-0.2 in}
\label{table:a}
\end{table*}
In the rest of the paper we compare \req{eq:alpha} with the experimental data from several high-mobility samples.
The 2DEG in each of our three samples (A, B, C) resides in a GaAs quantum well surrounded by Al$_{0.24}$Ga$_{0.76}$As barriers.
Electrons are supplied by Si doping in narrow GaAs wells, surrounded by thin AlAs layers, and placed at a setback distance $d$ on each side of the GaAs well hosting the 2DEG.
Sample parameters, such as density $n_e$, mobility $\mu$, quantum well width $w$, and setback distance $d$ are listed in Table~\ref{table:a}. 
The samples were $\approx 4-5$ mm squares with eight contacts positioned at the corners and at the midsides. 
Resistances $\Rxx$ and $\Ryy$ were measured using a four-terminal, low-frequency (a few Hz) lock-in technique at temperature $T \approx 50$ mK for sample A and at $T \approx 25$ mK for samples B and C.
The representative data for sample A are presented in \rfig{fig:data}.

For square sample geometry, resistivities $\rho_{xx}$ and $\rho_{yy}$ can be obtained from resistances $\Ryy$ and $\Rxx$ using \citep{simon:1999}
\begin{equation}
\label{eq:Simon}
R_{ii}=\frac{4}{\pi}\sqrt{\rho_{jj}\rho_{ii}}\sum_{n=\text{odd}^+}\left[n\sinh\left(\frac{\pi n}{2}\sqrt{\frac{\rho_{jj}}{\rho_{ii}}}\right)\right]^{-1}\,,
\end{equation}
where $i,j = x,y$ ($i \neq j$). 
This equation assumes that the current is passed between midside contacts and the voltage is measured between corner contacts.
With known $\Rxx$ and $\Ryy$, \req{eq:Simon} allows to obtain the resistivity ratio $\ar$ using
\begin{equation}
\label{eq:ar1}
\frac {\Rxx}{\Ryy} = \frac{\sum\limits_{n=\text{odd}^+}\left[n\sinh\left(\frac{\pi n}{2\sqrt{\ar}}\right)\right]^{-1}}{\sum\limits_{n=\text{odd}^+}\left[n\sinh\left(\frac{\pi n \sqrt{\ar}}{2}\right)\right]^{-1}}\,.
\end{equation}
Unfortunately, when $\Ryy$ becomes comparable to the experimental noise, direct application of \req{eq:ar1} becomes unreliable. 
In such situations we resort to using the parameter-free result for the resistivity product $\rho_{xx}\rho_{yy}$ which was first obtained by MF \citep{macdonald:2000} and, for half-integer $\nu$, can be written as
\begin{equation}\label{eq:product}
\rho_{xx}\rho_{yy}=\left(\frac{h}{e^2}\right)^2\frac{1}{(2\nu^2 + 1/2)^2}\,.
\end{equation}
This result, together with \req{eq:Simon}, allows one to obtain the resistivity anisotropy ratio from $\Rxx$ alone:

\begin{equation}
\label{eq:ar2}
R_{xx} = \frac {2h} {\pi e^2} \frac{\sum\limits_{n = \text{odd}^+}\left[n\sinh\left(\frac{\pi n}{2\sqrt{\ar}}\right)\right]^{-1}}{\nu^2 + 1/4}\,.
\end{equation}

\begin{figure}[b]
\vspace{-0.1 in}
\includegraphics{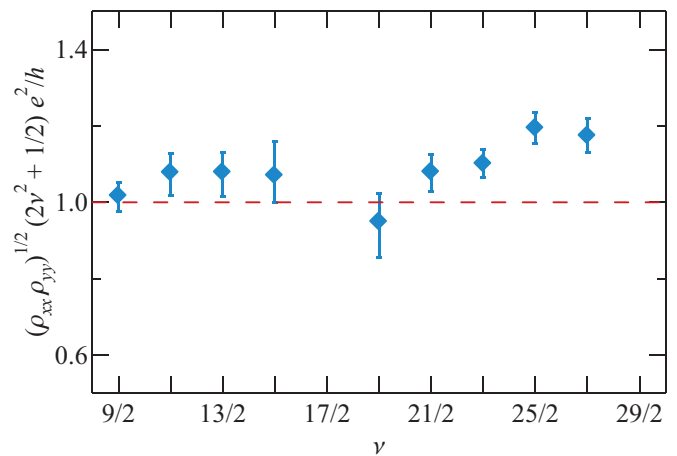}
\vspace{-0.1 in}
\caption{
(color online) $(\rho_{xx}\rho_{yy})^{1/2} (2\nu^2+1/2) e^2/h$ obtained using \req{eq:Simon} from $\Rxx$ and $\Ryy$ measured in sample B (diamonds) and prescribed by Eq.\,(\ref{eq:product}) (dashed line) vs. $\nu$.}
\vspace{-0.2 in}
\label{fig:product}
\end{figure}
While the validity of \req{eq:product} has been demonstrated long ago \citep{eisenstein:2001}, it is instructive to check it again.
In \rfig{fig:product} we present $(\rho_{xx}\rho_{yy})^{1/2}(2\nu^2 + 1/2)e^2/h$, obtained from \req{eq:Simon} using $\Rxx$ and $\Ryy$ measured in sample B, as a function of $\nu$ and observe that it stays close to unity as prescribed by Eq.~\eqref{eq:product} (dashed line).
Given the fact that \req{eq:product} has no adjustable parameters, the agreement is excellent and we will thus resort to using \req{eq:ar2} when $\Ryy$ cannot be reliably obtained.

\begin{figure}[t]
\includegraphics{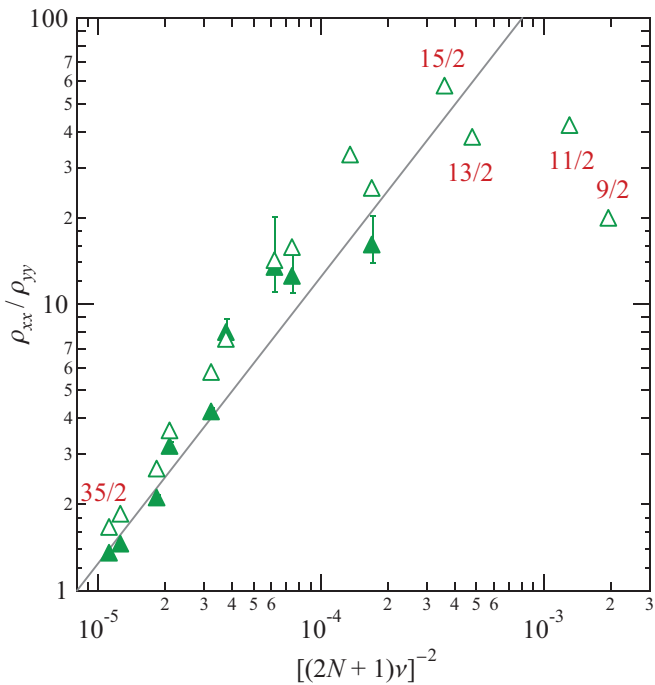}
\vspace{-0.1 in}
\caption{(color online)
Resistivity ratio $\ar$ in sample A obtained from \req{eq:ar1} (solid symbols) and \req{eq:ar2} (open symbols) as a function of $[(2N+1)\nu]^{-2}$. 
Half-integers mark filling factors.
The line represents \req{eq:alpha} with $\gamma = 0.15$. 
}
\vspace{-0.2 in}
\label{fig:a}
\end{figure}
We next compare our main theoretical result, \req{eq:alpha}, to the experimental resistivity ratio $\ar$ in sample A obtained using both methods.
In \rfig{fig:a} we present $\ar$ obtained from experimental resistances using  \req{eq:ar1} (filled triangles) and \req{eq:ar2} (open triangles) as a function of $[(2N+1)\nu]^{-2}$ over a wide range of half-integer $\nu$, $9/2 \le \nu \le 35/2$.
We observe that the values of $\ar$ obtained by different methods in general agree with each other.
The solid line is computed using \req{eq:alpha} with $\gamma =0.15$; it shows excellent agreement with the data for $13/2 \le \nu \le 35/2$. 
At $\nu \le 11/2$, however, we find that \req{eq:alpha} predicts $\ar$ which is considerably higher than what is observed in our experiment.
This trend is shared among all of our samples studied and we will return to this issue later.

\begin{figure}[t]
\includegraphics{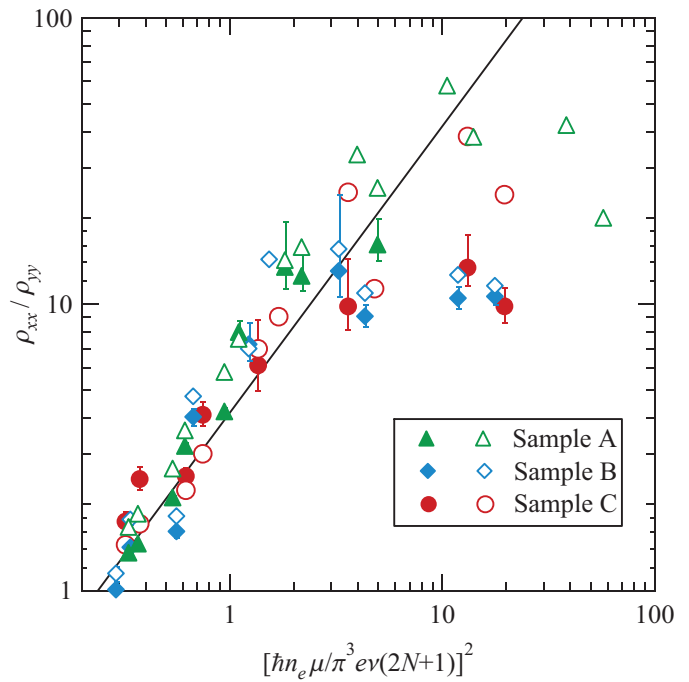}
\vspace{-0.1 in}
\caption{(color online)
Resistivity ratio $\ar$ in sample A (triangles), B (diamonds), and C (circles) obtained from \req{eq:ar1} (solid symbols) and \req{eq:ar2} (open symbols) as a function of a scaling variable $[\hbar n_e \mu/\pi^{3} e \nu(2N+1)]^2$. 
The line represents \req{eq:alpha} with $\gamma = 0.15$. 
}
\vspace{-0.2 in}
\label{fig:alpha}
\end{figure}
Having looked at the dependence of $\ar$ on $\nu$ and $N$, we next add its dependence on $n_e$ and $\mu$ in the scaling \rfig{fig:alpha}.
Here, we show the resistivity ratio $\ar$ obtained from \req{eq:ar1} (filled symbols) and \req{eq:ar2} (open symbols) for samples A, B, and C (see legend) as a function of $[\hbar n_e \mu/\pi^3 e \nu(2N+1)]^2 = [\hbar \sigma_0/\pi^3 e^2 \nu(2N+1)]^2$, where $\sigma_0$ is the conductivity at $B=0$. 
At $N \ge 3$, experimental points of \rfig{fig:alpha} are close to a line computed using Eq.~\eqref{eq:alpha} with $\gamma=0.15$.
For $N = 2$, however, we see substantial downward deviation of the data from this line. 
This is not surprising since we used \req{eq:E} with $\beta$ calculated for $N \gg 1$. 
Because $\ar \propto \beta^4$, only a 40$\%$ reduction of $\beta$ by quantum fluctuations would explain downward deviation of $N=2$ points.

Using $\gamma=0.15$ and \rfig{fig:gamma} we find that $N_1/N_2 \simeq 60$. 
This is in reasonable agreement with $N_1/N_2 \sim 10$ obtained previously in \rref{sammon:2018} from the analysis of mobility and quantum mobility of an ultrahigh mobility sample of similar design. 
As mentioned in \rref{sammon:2018}, large value of $N_1/N_2$ is likely related to a relatively impure Al source~\cite{chung:2018}. 
With $N_1/N_2 \simeq 60$ and scattering rates $1/\tau$ and $2/\tau_B$ calculated for both types of impurities in Supplemental Material [Eq.~(S1) and Eq.~(S2)], we can estimate concentrations $N_1$ and $N_2$ and their relative contributions to both rates. 
We find that for sample A, the spacer impurities have concentration $N_1 \simeq 5 \times 10^{14}$ cm$^{-3}$ and contribute 30$-$40\% to $2/\tau_B$ and 70$-$80\% to $1/\tau$, while GaAs well impurities have concentration $N_2 \simeq 10^{13}$ cm$^{-3}$ and contribute 60$-$70\% to $2/\tau_B$ and 20$-$30\% to $1/\tau$.

Obvious quantitative success of our \req{eq:alpha} to describe the record hard-to-easy resistivity ratios $\ar \lesssim 30$ for large range of parameters supports MF assumption that the low-temperature stripe phase is a pinned smectic phase predicted by HF calculations \citep{koulakov:1996,fogler:1996,moessner:1996}. Apparently, at least for $N \ge 3$, there is no evidence for the role of pinned or free dislocations and other large defects conjectured in Refs. \citep{fradkin:1999,fradkin:2000,oppen:2000}.
These conclusions agree with the recent observation of possible nematic-smectic transition at $T \lesssim 50$ mK~\cite{qian:2017}.

\begin{acknowledgements}
We thank I. Dmitriev and M. Fogler for discussions, and G. Jones, T. Murphy, and A. Bangura for technical support.
Calculations by M.S. and Y.H. were supported primarily by the NSF through the University of Minnesota MRSEC under Award No. DMR-1420013.
Experiments by X.F. and M.Z. were supported by the U.S. Department of Energy, Office of Science, Basic Energy Sciences, under Award \# ER 46640-SC0002567.
Growth of GaAs/AlGaAs quantum wells at Purdue University was supported by the U.S. Department of Energy, Office of Science, Basic Energy Sciences, under Award DE-SC0006671.
Growth of GaAs/AlGaAs quantum wells at Princeton University was supported by the Gordon and Betty Moore Foundation Grant No. GBMF 4420, and by the National Science Foundation MRSEC Grant No. DMR-1420541.
A portion of this work was performed at the National High Magnetic Field Laboratory, which is supported by National Science Foundation Cooperative Agreement Nos. DMR-1157490, DMR-1644779 and the State of Florida.
\end{acknowledgements}

\onecolumngrid
\section{\label{append}Supplementary material: calculation of $\gamma(N_1/N_2)$}
\renewcommand{\theequation}{S\arabic{equation}}

In order to find $\gamma(N_1/N_2)$ in Eqs.~\eqref{eq:alpha} and~\eqref{eq:tau_star}, we write both rates $1/\tau$ and $2/\tau_B$ as linear functions of the concentrations of background impurities in the spacer region ($N_1$) and in the quantum well ($N_2$):
\begin{align}\label{eq:tau_lin}
	&\frac{1}{\tau}=\frac{2\pi}{\hbar}\frac{g_0}{k_F}\left(\frac{2\pi e^2}{\kappa q_{TF}}\right)^2(a_1N_1+a_2N_2),\\
	&\frac{2}{\tau_B}=\frac{2\pi}{\hbar}\frac{g_B}{k_F}\left(\frac{2\pi e^2}{\kappa q_{TF}}\right)^2(b_1N_1+b_2N_2), \label{eq:tauB_lin}
\end{align} 
where $a_1$, $a_2$, $b_1$, and $b_2$ are dimensionless constants, $q_{TF} = 2/a_B$, and $k_F=\sqrt{2\pi n_e}$ is the Fermi wave number. 
Coefficients $a_{1,2}$ can be related to $A_{1,2}$ in the Eq.~(7) of the Ref.~\onlinecite{sammon:2018} as $A_{1,2} = a_{1,2}  k_F w \pi^2 \hbar/e$.
Following the definition of $\gamma(N_1/N_2)$ by Eq.~\eqref{eq:tau_star} and combining Eqs.\,(\ref{eq:tau_lin}) and (\ref{eq:tauB_lin}), we get
\begin{equation}\label{eq:gamma}
	\gamma(N_1/N_2)=\frac{b_1(N_1/N_2)+b_2}{a_1(N_1/N_2)+a_2}.
\end{equation}

We begin with the transport scattering rate in zero magnetic field
\begin{align}\label{eq:transport2}
	\frac{1}{\tau}=\frac{2\pi}{\hbar}g_0\frac{1}{2\pi}\int_0^{2\pi}\dd{\theta} \ev{\abs{V(q)}^2}(1-\cos{\theta}),
\end{align}
where $q=2k_F\sin(\theta/2)$, $\theta$ is the scattering angle, $\la \abs{V(q)}^2\ra$ is the Fourier transform of the scattering potential. Changing the integration variable from $\theta$ to $q$, we can write Eq.\,(\ref{eq:transport2}) as 
\begin{equation}\label{eq:transport3}
	\frac{1}{\tau} = \frac{4}{\hbar}g_0\int_0^{2k_F}\frac{\dd{q}}{k_F\sqrt{1-(q/2k_F)^2}}\left(\frac{q}{2k_F}\right)^2 \ev{\abs{V(q)}^2}.
\end{equation}
To proceed, we introduce the scattering potential for background impurities inside and outside of the well
\begin{equation}\label{eq:V(q)}
	\la \abs{V(q)}^2\ra=\left(\frac{2\pi e^2}{\kappa q\varepsilon(q)}\right)^2\left[N_1\frac{F_1(qw)}{q}+N_2\frac{F_2(qw)}{q}\right],
\end{equation}
where
\begin{equation}\label{eq:F_1}
	F_1(x)=\left[\frac{4\pi^2(1-e^{-x})}{x(4\pi^2+x^2)}\right]^2,
\end{equation} 
and
\begin{equation}\label{eq:F_2}
	F_2(x)=\frac{1}{x}\left(\frac{4\pi^2}{4\pi^2+x^2}\right)^2\left[\frac{8e^{-x}-e^{-2x}-7}{x}+2(2+e^{-x})+\frac{2x^2}{\pi^2}+\frac{3x^4}{8\pi^4}-\frac{8x(1-e^{-x})}{4\pi^2+x^2}\right],
\end{equation}
are form factors of the quantum well wave function~\cite{sammon:2018}.
We have also introduced the dielectric screening function
\begin{equation}\label{eq:epsilon}
	\varepsilon(q)=1+\frac{q_{TF}}{q}G(qw),
\end{equation}
with the form factor~\cite{sammon:2018}
\begin{equation}\label{eq:G(x)}
	G(x)=\frac{20\pi^2x^3+3x^5-32\pi^4(1-x-e^{-x})}{x^2(4\pi^2+x^2)^2}.
\end{equation}
Combining Eqs.\,(\ref{eq:transport3})-(\ref{eq:G(x)}), it is easy to see that 
\begin{equation}\label{eq:A1}
	a_{1,2}=\frac{2}{\pi}\int_0^{2k_F}\frac{\dd{q}}{q\sqrt{1-(q/2k_F)^2}}\frac{F_{1,2}(qw)}{(q/q_{TF}+G(qw))^2}\left(\frac{q}{2k_F}\right)^2.
\end{equation}
For $n_e = 3\times 10^{11}$ cm$^{-3}$ and $w = 30$ nm, we find $a_1=0.011$ and $a_2=0.20$.

Now we calculate the rate $2/\tau_B$ for an electron on the edge of a stripe to scatter onto an adjacent edge. We begin with Fermi's golden rule for the scattering rate from an initial state $\ket{i}$:
\begin{equation}\label{eq:Fermi}
	\frac{2}{\tau_B}=\frac{2\pi}{\hbar}\sum_f \ev{\abs{\mel{f}{\hat{V}}{i}}^2} \delta(\varepsilon_f-\varepsilon_i),
\end{equation}
where $\hat{V}$ is the operator for the disorder potential $V(\vb{r})$, $\ev{\dots}$ denotes averaging over disorder realizations, and the summation is over all the final states $\ket{f}$. The matrix element can generally be written as 
\begin{equation}\label{eq:matrix_element}
	\ev{\abs{\bra{f}\hat{V}\ket{i}}^2} = \int \dd[2]{r} \dd[2]{r'}\psi_{f}^*(\vb{r})\psi_{i}(\vb{r})\psi_{f}(\vb{r'})\psi_{i}^*(\vb{r'}) \ev{ V(\vb{r})V(\vb{r'})},
\end{equation}
where
\begin{equation}\label{eq:potential_corr}
	\ev{V(\vb{r})V(\vb{r'})}=\int \frac{\dd[2]{q}}{(2\pi)^2}\ev{\abs{V(q)}^2} e^{-i\vb{q}\vdot(\vb{r}-\vb{r'})},
\end{equation}
and $\ev{\abs{V(q)}^2}$ is defined in Eq.~\eqref{eq:V(q)}. Substituting Eq.\eqref{eq:potential_corr} into Eq.~\eqref{eq:matrix_element} we arrive at 
	\begin{equation}\label{eq:matrix_element2}
		\ev{\abs{\mel{f}{\hat{V}}{i}}^2} = \int\frac{\dd[2]{q}}{(2\pi)^2}\ev{\abs{V(q)}^2} \abs{\int \dd[2]{r}\psi_{f}^*(\vb{r})e^{-i\vb{q}\vdot\vb{r}}\psi_{i}(\vb{r})}^2.
	\end{equation}	
In the Landau gauge, $\vb{A}=Bx\vu{j}$, the wave functions of the $N$th LL can be written as
\begin{gather}\label{eq:wavefunction}
	\psi_{X_{i,f}}^N(\vb{r})=\frac{e^{iyX_{i,f}/ l_B^2}}{\sqrt{L_y}}\chi_N(x-X_{i,f}),\\
	\chi_N(x)=\frac{1}{\pi^{1/4}\sqrt{2^NN! l_B }}\exp\left[-\frac{x^2}{2 l_B^2}\right]H_N\left(\frac{x}{ l_B}\right),\label{eq:chi_xn}
\end{gather} 
where $X_{i}$ and $X_{f}$ are the $x$-coordinates of the cyclotron center of the initial and final states respectively, and $L_y$ is the sample length in the $y$-direction. We can assume without loss of generality that $X_i=0$ and find
\begin{equation}\label{eq:FT_Wavefunction}
	\int \dd[2]{r}\psi_{f}^*(\vb{r})e^{-i\vb{q}\vdot\vb{r}}\psi_{i}(\vb{r})=\frac{2\pi}{L_y}\delta\left(q_y-\frac{X_f }{ l_B^2}\right)\exp\left(\frac{iq_x X_f}{2}\right)\Phi_N(q l_B),
\end{equation}
where 
\begin{equation}\label{eq:FormFactor}
	\Phi_N(q l_B)=\exp\left(-\frac{q^2 l_B^2}{4}\right)L_N\left(\frac{q^2 l_B^2}{2}\right),
\end{equation}
and $L_N(x)$ is the $N$th Laguerre polynomial. 
Substituting Eq.\,(\ref{eq:FT_Wavefunction}) into Eq.\,(\ref{eq:matrix_element2}) and performing the integrals over $y$ and $q_y$ we get
\begin{equation}\label{eq:matrix_element3}
	\ev{\abs{\mel{f}{\hat{V}}{i}}^2} = \frac{1}{L_y}\int_{-\infty}^{\infty}\frac{\dd{q_x}}{2\pi} \ev{\abs{V(q)}^2} \Phi_N^2(q l_B),
\end{equation} 
where $q^2=q_x^2+X_f^2/ l_B^4$. Returning to Eq.\,(\ref{eq:Fermi}), we write the summation over final states as 
\begin{equation}\label{eq:summation}
	\sum_f(...)\delta(\varepsilon_i-\varepsilon_f)=\frac{L_y}{2\pi l_B^2eE}\int \dd{X_f}(...) [\delta(X_f-\Lambda/2)+\delta(X_f+\Lambda/2)].
\end{equation}
Combining this with Eq.\,(\ref{eq:matrix_element3}), performing the integral over $X_f$, we can finally write
\begin{equation}\label{eq:tau_B}
	\frac{2}{\tau_B}=\frac{2\pi}{\hbar}g_B\Lambda\int_{-\infty}^{\infty}\frac{\dd{q_x}}{2\pi}\ev{\abs{V(q)}^2} \Phi_N^2(q l_B).
\end{equation}
Substituting the potential defined in Eq.\,(\ref{eq:V(q)}), we arrive at
\begin{equation}\label{eq:B1}
	b_{1,2}=\frac{k_F\Lambda}{\pi}\int_0^{\infty}\frac{\dd{q_x}}{q}\Phi_N^2(q l_B)\frac{F_{1,2}(qw)}{(q/q_{TF}+G(qw))^2},
\end{equation}		
In Eqs.~(\ref{eq:tau_B}-\ref{eq:B1}), $q^2 = q_x^2 + \Lambda^2/4 l_B^4$.
Following Ref.~\onlinecite{wexler:2001} we choose $\Lambda = 2.84 R_c$.
For $N<25$ studied numerically, the coefficients $b_1(N)$ and $b_2(N)$ oscillate with period $\Delta N \simeq 5$ and amplitude $\sim 50\%$ around their averages as a result of the oscillation of wave functions Eq.~\eqref{eq:wavefunction}. 
These oscillations can be substantially reduced due to fluctuations of charge density of the remote doping layers, which induce fluctuations $\Delta \nu$ of the 2DEG filling factor around half integer $\nu$.
To estimate $\Delta\nu$ we used results of numerical modelling of doping layers of modern GaAs/AlGaAs devices~\cite{sammon:2018b}, which showed that in the ground state of localized excess electrons of the doping layers with fraction of filled donors $f\simeq 0.5$, the mean square fluctuation of charge of the square of size $L$ around average value is $\sim 1e$ practically independent on $L$. 
Then in area $\Lambda^2$ one has $\Delta\nu \simeq 2^{1/2} 2\pi  l_B^2 / \Lambda^2 \simeq 1/(2N+1)$, where the factor $2^{1/2}$ results from the addition of mean square fluctuations of the two doping layers.
Such fluctuations produce fluctuations of the width of filled and empty stripes $\Delta \Lambda \simeq 1.42 R_c /(2N+1)$, which reduce the amplitude of oscillations of $b_{1,2}(N)$ at least twice and allow us to use their average values $b_1 = 7.5 \times 10^{-4}$  and $b_2 = 8.9 \times 10^{-2}$. 
Substituting these values of $b_1$ and $b_2$ together with $a_1$ and $a_2$ found above into Eq.~\eqref{eq:gamma} we arrive at Fig.~\ref{fig:gamma} for the function $\gamma(N_1/N_2)$ used in the main text.


\end{document}